# Nanomechanical morphology of amorphous, transition, and crystalline domains in phase change memory thin films

J. L. Bosse[1], I. Grishin[2], B. D. Huey[1], O. V. Kolosov[2,a]

[1]Department of Materials Science & Engineering, University of Connecticut, 97 North Eagleville Road, Unit 3136, Storrs, CT 06269-3136
[2]Department of Physics, Lancaster University, Lancaster, United Kingdom, LA1 4YB

a) Electronic mail: o.kolosov@lancaster.ac.uk, Phone: +44 (0)1524 593619

**Abstract**

*In the search for phase change materials (PCM) that may rival traditional random access memory, a complete understanding of the amorphous to crystalline phase transition is required. For the well-known $Ge_2Sb_2Te_5$ (GST) and GeTe (GT) chalcogenides, which display nucleation and growth dominated crystallization kinetics, respectively, this work explores the nanomechanical morphology of amorphous and crystalline phases in 50 nm thin films. Subjecting these PCM specimens to a lateral thermal gradient spanning the crystallization temperature allows for a detailed morphological investigation. Surface and depth-dependent analyses of the resulting amorphous, transition and crystalline regions are achieved with shallow angle cross-sections, uniquely implemented with beam exit Ar ion polishing. To resolve the distinct phases, ultrasonic force microscopy (UFM) with simultaneous topography is implemented revealing a relative stiffness contrast between the amorphous and crystalline phases of 14% for the free film surface and 20% for the cross-sectioned surface. Nucleation is observed to occur preferentially at the PCM-substrate and free film interface for both GST and GT, while fine subsurface structures are found to be sputtering direction dependent. Combining surface and cross-section nanomechanical mapping in this manner allows 3D analysis of microstructure and defects with nanoscale lateral and depth resolution, applicable to a wide range of materials characterization studies where the detection of subtle variations in elastic modulus or stiffness are required.*

## 1.0 Introduction

Significant efforts continue to try to improve non-volatile memory systems, ideally with improved read/write cycle endurance, faster switching speeds, and lower power consumption. One such class of materials are the ternary chalcogenides, which exhibit rapid and reversible phase transitions between the amorphous and crystalline states [1, 2]. It is well known that this class of phase change materials (PCM) [3] still needs to be improved in several key areas to become competitive with current technologies, with the major target being finding a stoichiometry that exhibits a fast crystallization speed [4] while maintaining mechanical and morphological stability upon high cycling of the read and write process [5]. In parallel with experimental [6-11] and theoretical studies [12-14], nanoscale characterization methods such as scanning probe microscopy (SPM) [8, 15-18] and transmission electron microscopy (TEM) [10, 12, 19, 20] have been widely implemented to study the phase switching dynamics and mechanical properties of the chalcogenide materials.

One particularly useful tool for studying the nanomechanical morphology of the switched phases and their corresponding stresses due to density changes combines nanomechanical mapping by ultrasonic force microscopy (UFM) with beam exit Ar ion beam polishing (BEXP) [21] for low-damage, shallow angle cross sectioning. Accordingly, we report the 3-dimensional nanomechanical morphology of amorphous and crystalline phases for two commercially viable phase change stoichiometries, $Ge_2Sb_2Te_5$ (GST) and GeTe (GT), each with thermally and optically induced crystallization. This work is particularly relevant for characterizing defects through the film thickness for ultimately improving device design and dimensional scaling of these phase change technologies.

## 2.0 Methods

### 2.1 Sample Fabrication

Amorphous GeTe and $Ge_2Sb_2Te_5$ films were sputtered (Moorfield MiniLab 25) onto Si wafers (3" diameter, 280 μm thick, p-doped, <100>, 0.01 – 1.0 Ω-cm), with an intermediate 100 nm of sputtered Ti to promote adequate bonding. The sputtering conditions in all cases include a deposition rate of 0.3 – 0.4 Å/s at a base pressure of $10^{-5}$ Torr, with an RF power of 6-8 W for both chalcogenide layers and 40 W for the Ti matching layer.

### 2.2 Sample Processing

Following sample fabrication, each wafer was cut into 25 x 75 mm strips for the application of a thermal gradient, thereby nucleating the crystalline phase in the amorphous film with a distinct transition region (containing both amorphous and crystalline phases) in between. One side of each strip was attached to an electric heater capable of controlled heating to 600 $^0$C (Linkam Scientific Instruments, UK), while the other side was mounted 20 mm away to a carbon steel heat sink (20 mm diameter, 3.2 mm thickness), Figure 1(a).

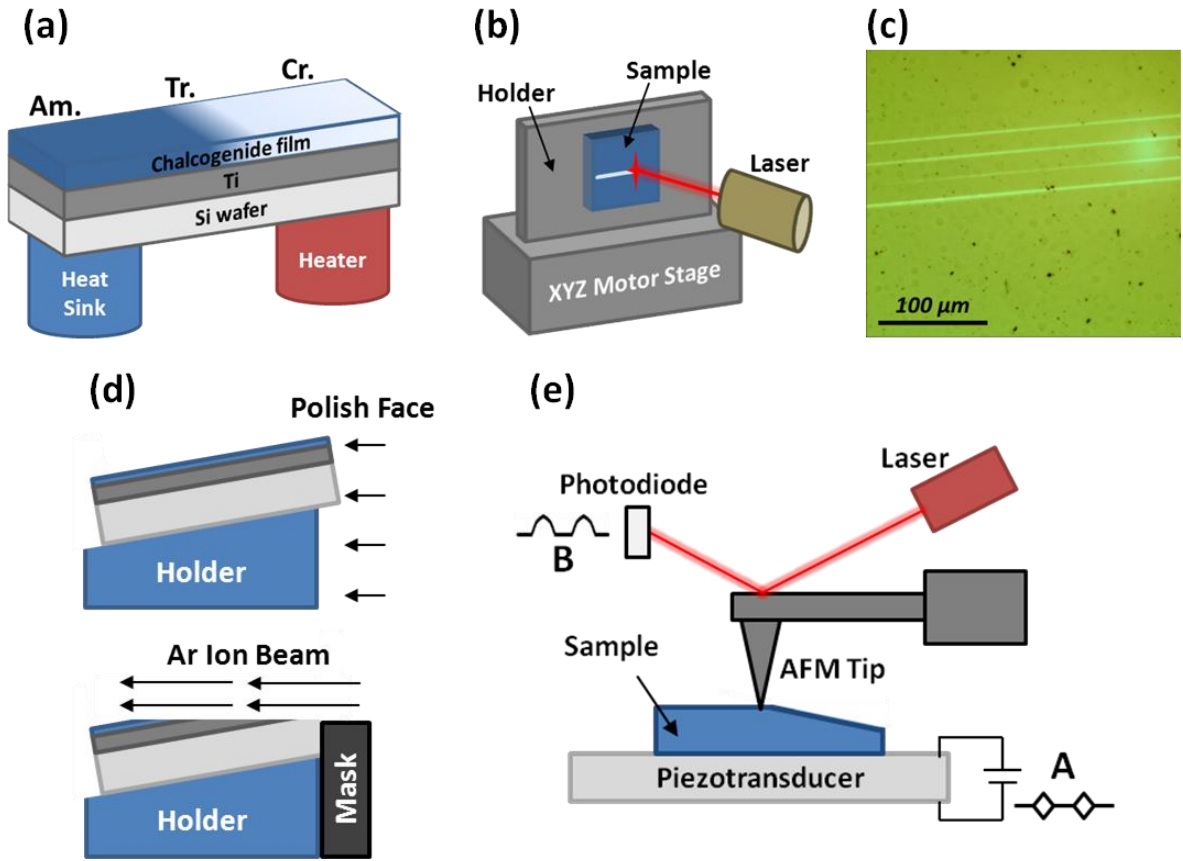

**Figure 1:** As deposited predominantly amorphous phase change material samples were mounted onto a heater and heat sink via thermally conductive paste creating a fully crystalline phase at the heater side, unaffected amorphous area and a transition region (a). All samples with distinct regions (amorphous, transition, crystalline) were mounted onto an XYZ motor stage and locally heated with a CW laser providing a narrow line of fully switched crystalline zone (b), with phase switching then confirmed by optical microscope (c). The samples were then cross-sectioned using BEXP method [21] (d) for nanomechanical characterization by UFM [22] (e).

Adequate thermal contact between both the sample/heater and sample/heat sink surfaces was made by Boron Nitride conductive heat sink grease (CircuitWorks CW7250). The temperature of the heater was increased by 10°C/s from room temperature to 300°C and held for 2 minutes, until complete crystallization occurred for the phase change material directly above the heater surface. The heat sink was present to maintain the amorphous region in one end of the film (although its temperature was not monitored, no notable heating occurred due to the large mass of the heat sink and small size of the PCM sample). The transition regions between amorphous and crystalline states for GT and GST films were 350-450 μm and 900-950 μm wide, respectively, as confirmed by optical microscopy.

Following the thermal gradient treatment, each strip of GT and GST was cleaved into three 6 x 6 mm samples, separating the amorphous, crystalline, and transition regions. Each sample was mounted onto a motorized XYZ stage and a focused CW laser of 30-40 mW on-the-sample power (514 nm Ar ion laser, Spectra Physics, USA) was used to write an optically-induced crystalline 'reference' line (2-5 µm wide, 2-3 mm long) across all three phase regions, Figure 1(b). The samples were programmatically translated with a step motor controller (Honda Electronics, Japan) at 50 µm per second to provide a consistent Joule heating per unit area as described elsewhere [11]. Further characterization under the optical microscope confirmed contrast between the crystalline reference line and the amorphous/transition regions for GT and GST, but no observable contrast in the already-crystalline region as expected. This both confirms the complete transition of the thermally switched area, as well as the absence of possible artefacts from laser heating alone. Figure 1(c) typifies 4 parallel crystalline phase lines written in the amorphous GST film as observed in the optical microscope.

Cross-sectioning of all three phase regions for each PCM stoichiometry was next performed by the BEXP method, Figure 1(d), using an in-house modified cross-section polisher (Leica EM TIC020, Germany) [23]. To achieve a final polished angle of 10° from the surface normal, the side surface of each sample was mechanically lapped with a 80° angle to ensure adequate contact with the mask. The ion polisher cutting voltage was set to 7 kV, until the cross-section was completed. The cutting voltage was then lowered to 1 kV for 15-30 minutes to finely polish the cross-section and prevent sample erosion due to transmission sputtering [24].

*2.3    Sample Characterization*

In order to quantify the nanoscale morphology and relative stiffness of the chalcogenide phases, scanning probe based methods were implemented. Specifically, UFM was chosen due to its ability to distinguish subtle differences in local elasticity (<0.1%) [25] for even the most stiff materials, and with the same or better lateral resolution as conventional contact mode AFM [26-28]. Figure 1(e) displays the schematic for these UFM measurements, allowing investigation of both the free surface and the shallow-angle cross sections in a single image. All UFM measurements were performed on both a Multimode SPM system with Nanoscope IV controller (Bruker, USA) and a Cypher AFM (Asylum Research, USA) system, each operated in an ambient environment. To maximize the propagation of the longitudinal ultrasonic waves key to the UFM technique, all samples were mounted onto a piezotransducer with crystalline salol (phenyl salicylate, melting point 42°C). Ultrasonic vibrations were excited at the resonant frequency of the 4 MHz piezotransducer (Physik Instrumente, Germany) with an amplitude modulated sine wave defined by a triangular-shaped waveform produced by an arbitrary function generator (Agilent 33220A, USA), signal A in Figure 1(e). The frequency of the arbitrary waveform was chosen (2.71 kHz) as an optimal trade-off, avoiding vertical feedback loop influences and cantilever resonances, while allowing for sufficiently fast data acquisition. In response to these periodic ultrasonic vibrations, correlated periodic normal displacement of the AFM tip occurs detected by the AFM's position sensitive quadrant photodiode (signal B in Figure 1(e)). This normal deflection signal was analysed by a lock-in amplifier (LIA) (Stanford Research Systems 830), with the reference signal at the modulation frequency provided by the external waveform generator. The UFM amplitude $U_z$ at the first harmonic of the modulation frequency was finally extracted, with a filter time constant of 1 to 3 ms for proper averaging up to the duration of each image pixel given line scan rates usually of 0.5 to 1.0 Hz. Typical LIA sensitivities were 10 to 20 mV full scale, appropriate for ultrasonic carrier sine wave amplitudes in the range of 2 to 5 $V_{PP}$. The UFM amplitude was finally routed to the AFM auxiliary inputs, allowing simultaneous topography and nanomechanical UFM acquisition.

*2.4   Image Analysis*

A standard fast Fourier transform (FFT) band pass filter was applied to each UFM image to remove periodic noise, with the lower limit of the filter set to 50% of the grain size, approximately 30 nm for all regions of GST and 15 nm for GT, respectively.

## 3.0    Results and Discussion

### 3.1    *Plan View Specimens*

Prior to performing a complete nanomechanical characterization (cross-sectioning and imaging), the nature of each amorphous, transition, or crystalline phase region was verified by two methods. First, the CW laser was used to write a well-defined crystalline-phase-only reference line (see Methods). As these PCM materials display a high reflectivity contrast between amorphous and crystalline phases, allowing their easy identification [29], the films were then inspected with optical microscopy. The films were then independently characterized by UFM, allowing the observation of features much finer than wavelength-limited optical techniques. The crystalline line present in all samples studied also serves as a consistent reference for UFM measurements up to 10-20 μm on either side. This alleviates the need for independent calibration of the absolute ultrasonic amplitude (and hence UFM contrast) that is otherwise necessary, since it can vary on the scale of many hundreds of μm [30] due to the underlying transducer, resonant effects, or spatially dependent ultrasonic transmission. Since the μm scale UFM images presented here are orders of magnitude smaller than the ultrasonic wavelength, and include an identical crystalline region in each case, they may thus be compared with confidence [31].

Figure 2 presents representative topography (left) and UFM (right) images for the amorphous and transition phase regions for the GST and GT films. Images for the crystalline reference line through the crystalline phase region for GST and GT were also acquired, but are not presented as they display no topography or ultrasonic contrast as expected. The location of the crystalline reference line (dashed lines on topography) is apparent in the topography images by a depression in the thin film, as a result of the lower specific volume for the crystalline phase [32]. The UFM images of the crystalline line through the amorphous region (top row) displays a stronger (brighter) UFM response for the crystalline versus amorphous phase, indicating a higher relative stiffness [33]. This behavior is expected based on previous mechanical studies by nanoindentation [34, 35], wafer curvature [36], and resonating beam methods [37].

The topography and UFM images of the crystalline lines through the transition regions (bottom row) also reveal an increased density and stiffness, respectively, but with less contrast presumably due to the partial phase change that has already occurred. The laser line passing through the transition region in GST displays particularly interesting topographic and nanomechanical morphology. The center of the laser line has a lower topography with corresponding high contrast, indicative of crystalline material. Approximately 1-2 µm to either side of the laser line center, however, both the topography and ultrasonic response suggest a higher amorphous fraction. This behaviour may reflect a complex strain distribution in such nanometre scale thick films, especially where the laser induced crystallization is performed in an area already possessing crystalline nuclei as expected for the thermally induced transition region. Another possibility is the film may partially delaminate from the substrate so that a lower ultrasonic response would be locally observed [38].

For the thermally induced transition region, but away from the crystalline reference lines (i.e. not optically switched), a profound difference in crystallization morphology is apparent for the GST versus GT. For the GST film the microstructure is too fine to be properly resolved in these ~8 µm images, but is clearly less than 100 nm. For GT, on the other hand, there are clear, circular shaped, 2- 5 µm diameter crystalline regions.

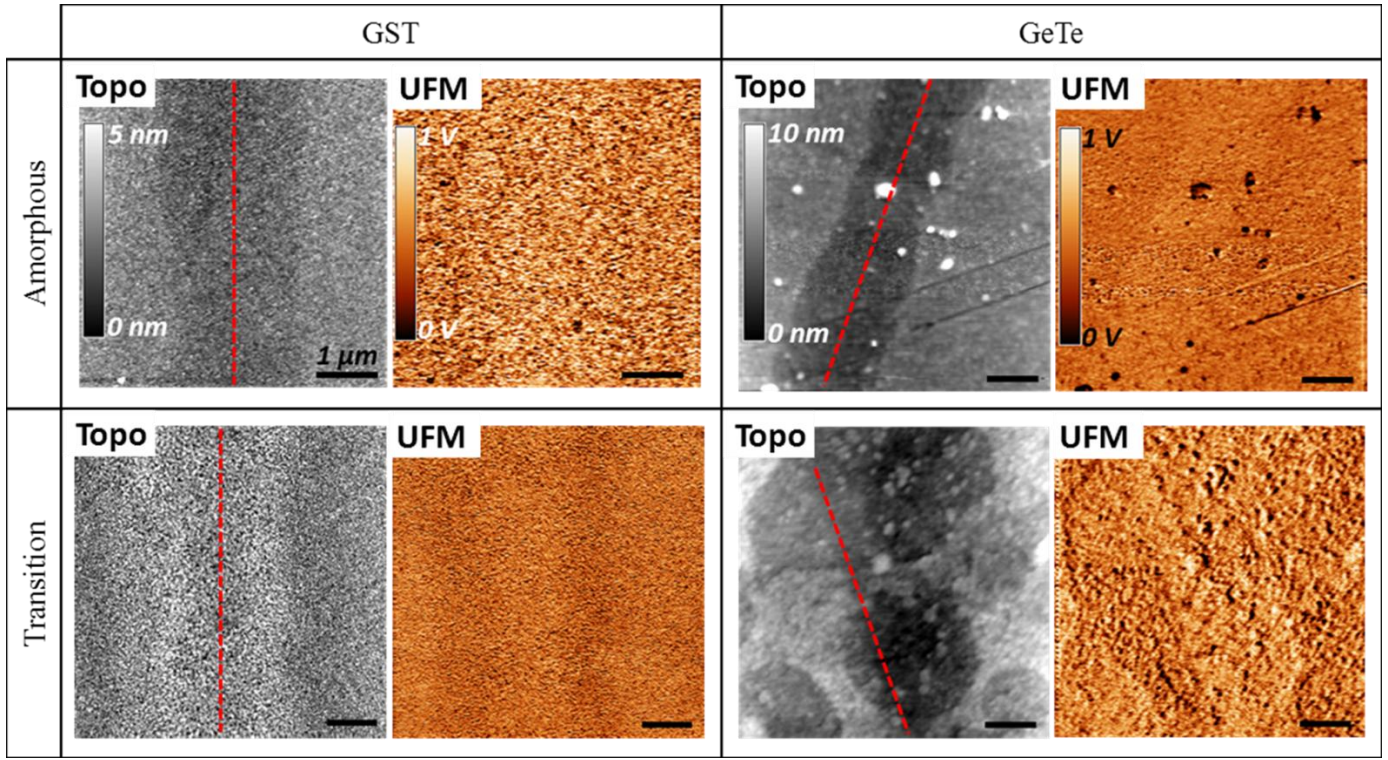

**Figure 2:** Topography (1st and 3rd columns) and UFM images (2nd and 4th columns) of amorphous (top row) and transition (bottom row) regions of GST (left) and GT (right) films with a crystalline reference line written by laser pulses according to Figure 1(b). The centreline of the laser path is identified in each topography image (dashed overlay), with uniform 1 μm scale bars shown throughout.

To explore details of the nanomechanical morphology of individual grains, higher magnification (1 μm) images were acquired away from but still sufficiently close to the crystalline reference line for reliable UFM measurements, Figure 3. For both the amorphous and crystalline phase regions of the GST film, the grains were ellipsoidal with height and width of ~55 and ~70 nm, respectively, in close agreement with the ~70 nm grain size reported for similarly sputtered GST films [39]. One notable feature in the UFM images for the GST crystalline phase is the appearance of dark, less stiff rings surrounding some grains (see 3x magnification insets). These may be amorphous residues that fail to crystallize at the grain boundaries, with a characteristic size of ~6 nm as reported elsewhere [40, 41]. There are occasional similar features present within the amorphous region as well, which could result from the amorphous region of GST containing a small fraction of crystalline phase grains, likely due to localized temperature elevations during the initial sputtering process.

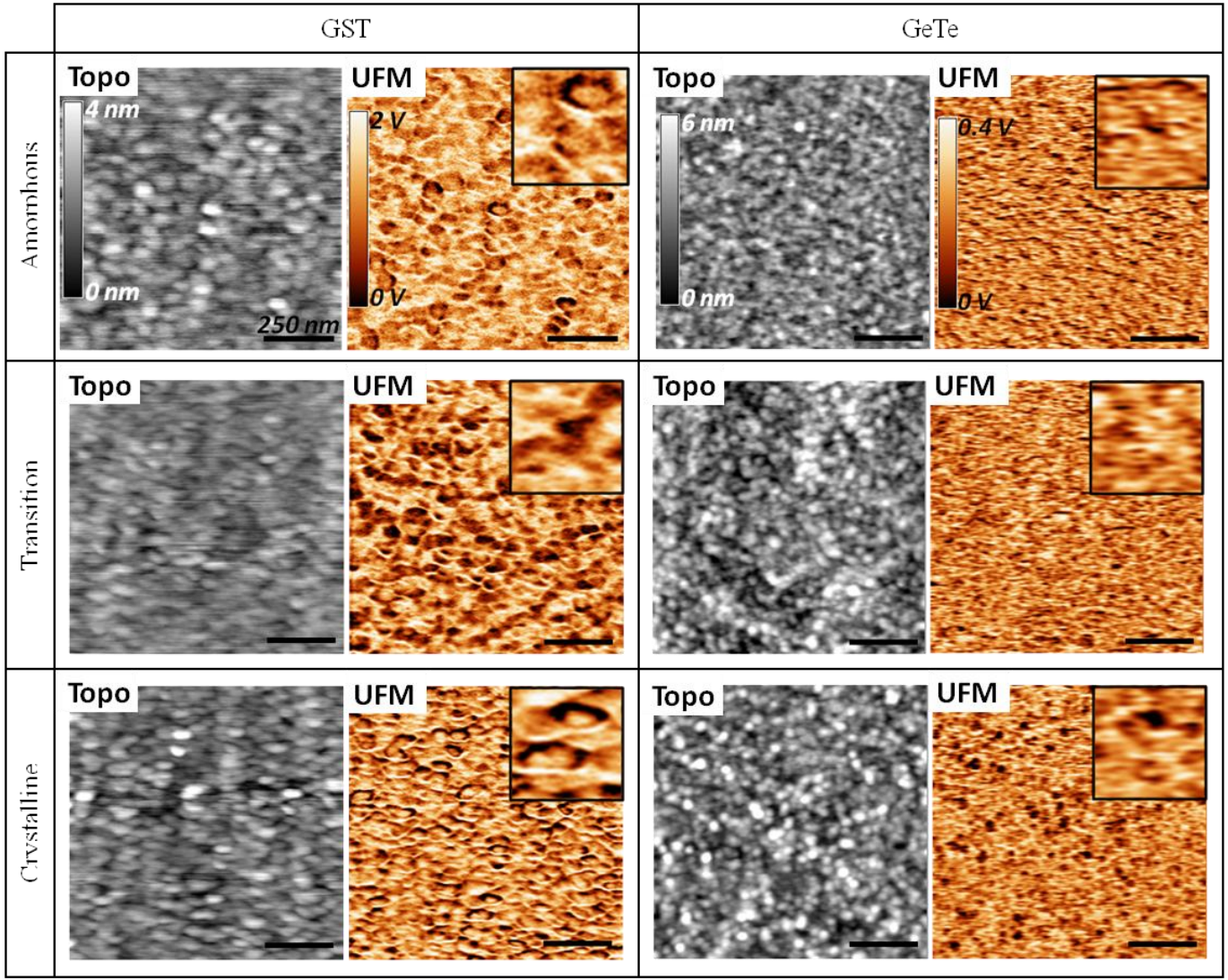

**Figure 3: 1 μm topography (left) and UFM images (right) of amorphous (top row), transition (center row), and crystalline (bottom row) regions of GST and GT films. Insets are provided within each UFM image (3x magnification), clearly resolving the nanomechanical morphology.**

The transition region for GST (middle row in Fig. 3) has much higher nanomechanical non-uniformity than the amorphous or crystalline regions, revealing interspersed amorphous/crystalline phases with varied stiffness. The characteristic length scale of the nucleated crystalline phase within the transition region is approximately 100 – 200 nm, similar to crystallization behavior reported by Yang et al. [15]

To more accurately compare the relative stiffness of all three phase regions, which are necessarily obtained over different regions of the underlying piezotransducer, the RMS variation of the UFM signal was normalised to the average response over the entire imaged area as described elsewhere [30]. For GST, the UFM stiffness contrast in the thermally induced amorphous and crystalline regions varied by 7% and 11%, respectively. For the transition region, it varied significantly more strongly at 14%. On the other hand, the amorphous, crystalline, and transition regions uniformly displayed a relatively weak ultrasonic response variation of 3%, 4%, and 2%, respectively, possibly reflecting the difference between the nucleation dominated kinetics of GST vs. growth dominated kinetics of GT.

*2.1    Shallow Angle Cross Sections*

Following these unique measurements of the surface nanomechanics of variously switched films, BEXP prepared shallow-angle cross-sections for each phase region of GST (top row) and GT (bottom row) were also studied with UFM (Figure 4). Every image has been rotated for clarity, such that the top portion of each image displays the intact surface of the phase change material (akin to the results of Figures 2 and 3, exemplified by the region labelled 'A' in Figure 4). The next layer from the top is the 50 nm cross-sectioned phase change material, followed by 100 nm of cross-sectioned Ti, and the underlying cross-sectioned Si substrate.

Importantly, the ion-cross-sectioned portions of the presented images (i.e. beyond region A) are at an oblique orientation topographically, as identified in Figure 1(e,f). Therefore, the stated thickness of each layer is not directly apparent. Moreover, due to variations in the speed at which the Ar ions penetrate the film as well as the local sputtering yield [21], the sectioned angle differs slightly from one layer to another. Conveniently, the simultaneously acquired topography line profiles (Figure 5) can identify these angles, ranging from 17 to 40 degrees, and can be used to calibrate the actual layer thicknesses in the oblique view.

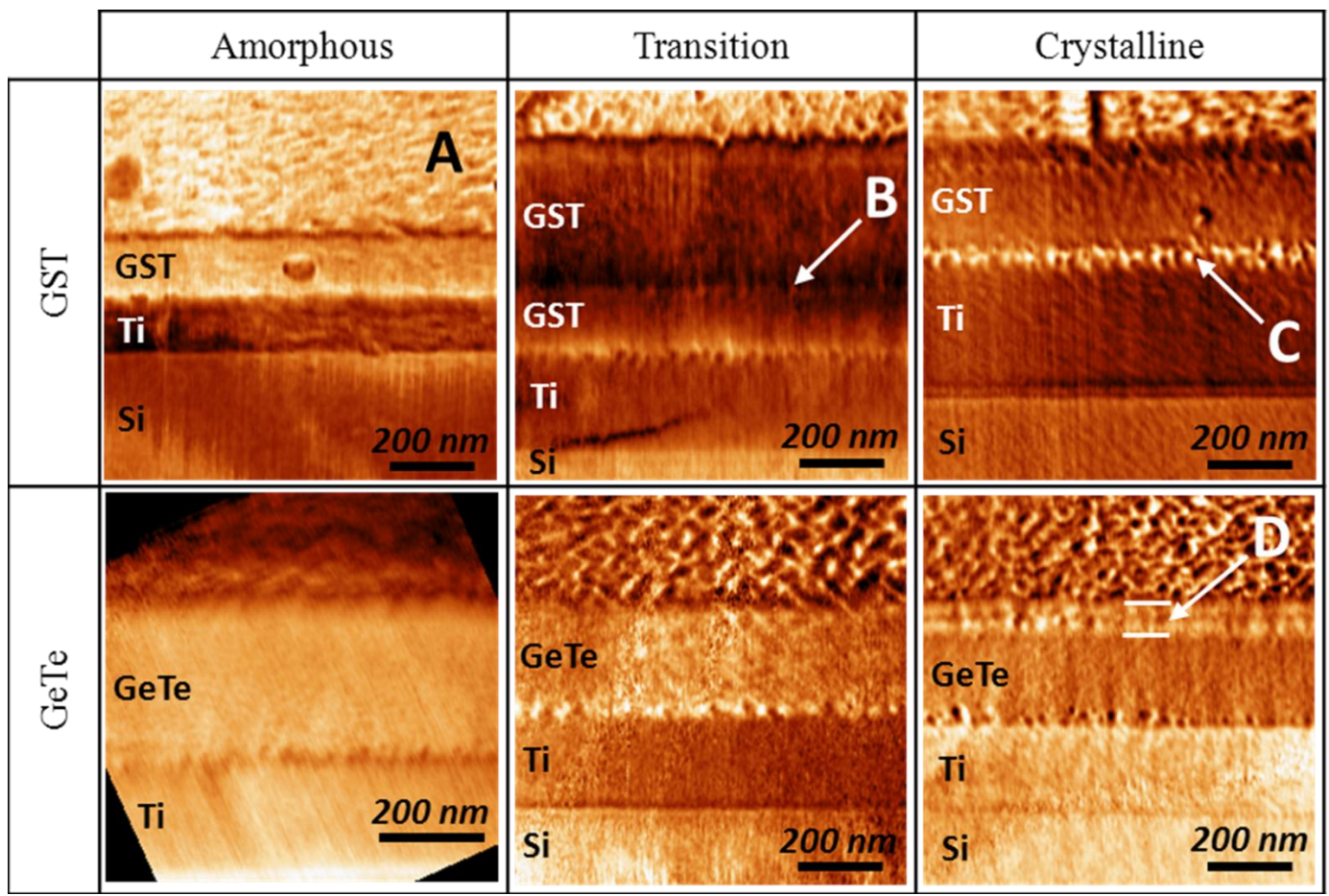

**Figure 4:** UFM Images of GST and GT films with BEXP cross-sectioning performed according to Figure 1(d). Note: images are presented at an oblique orientation, therefore the thickness of phase change material (50 nm) and Ti underlayer (100 nm) are not directly apparent without topography line profiles (Figure 5).

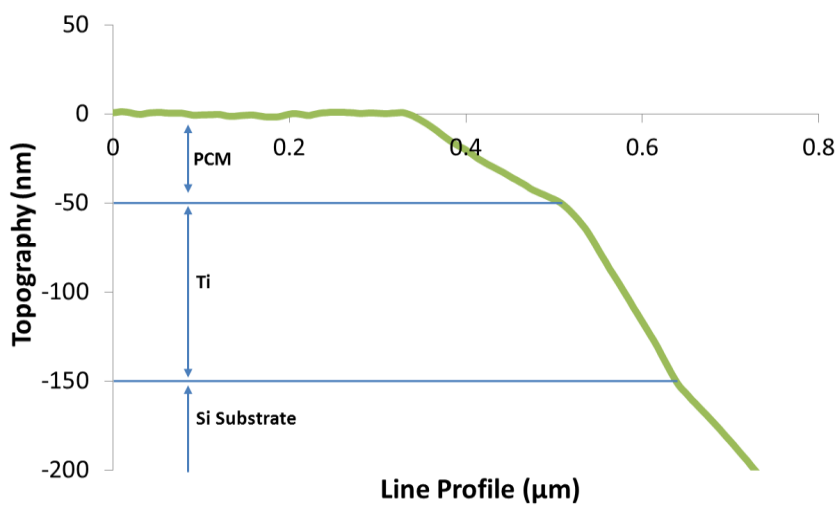

**Figure 5:** Example topography line-scan profile for the cross-sectioned phase change material, taken from the GST amorphous region, line 66 (out of 256) from the top. The angles between the free-film and cross-sectioned surfaces are 17°, 40°, and 29° for PCM, Ti, and the Si substrate, respectively.

First considering the topography of the sections, rms-roughness values for the amorphous, transition, and crystalline regions of GST are 0.6 nm, 1.4 nm, and 1.4 nm, respectively. For GT, the amorphous, transition, and crystalline regions are 1.1 nm, 0.5 nm, and 1.0 nm, respectively. The corresponding normalized rms calculations of the UFM signal variability, as discussed for Figure 3, is summarized in Table I, revealing differences between the UFM response for the top (normal to the substrate) and cross-sectioned surfaces. This is likely due to the slight topographic variations from one region to the other, which could otherwise strongly influence contact-nanomechanical results due to changes in contact area [42]. This highlights a benefit of cross-sectioning with the BEXP technique, as it can reduce such topographic influences on nanomechanical contrast by generating surfaces with roughness on the order of just 1 nm as occurred here [23].

With respect to the UFM contrast, there are several noteworthy observations. First, the GST film in the transition region contains a discontinuity (labelled 'B' in Figure 4) which displays a lower ultrasonic response. We believe this corresponds to the sputtering process, which was interrupted part way through deposition of the GST layer, highlighting the sensitivity of the current approach. Additionally, the ultrasonic response at this interface is weaker than the surroundings, indicating a decrease in the local stiffness. Second, the ultrasonic response at the junction between the phase change layers and the Ti (labelled 'C' in Figure 4) displays a stronger UFM contrast for the crystalline and transition regions, but not for the amorphous regions. Such higher signals at the PCM/Ti interface suggest locally stiffer materials as a result of a nucleated crystalline phase. However, a heterogeneity due to possible formation of a $TiO_x$ layer cannot be excluded as a result of breaking the vacuum while switching the sputtering target material from Ti to GST/GT. In future, the sample processing may be improved by sputtering within a chamber capable of holding multiple source materials, to ensure that the vacuum is maintained.

Third, at the interface between the top surface and cross-sectioned phase change material for the crystalline GST/GT and GST transition (exemplified by 'D' in Figure 4), a 5-10 nm layer with stronger nanomechanical contrast also is apparent. This most likely is a known artefact associated with the BEXP process near the original specimen surface, where if the Ar beam is allowed to continue after the cross-section is completed, transmission sputtering through the top surface will erode the surface material [23]. This may be addressed by lowering the Ar beam power from 5-7 kV down to 1 kV at the exact moment the beam finishes cutting, but this is difficult to implement without an automated cutting process. Nevertheless, the ability to characterize these artefacts with UFM demonstrates how powerful a tool nanomechanical mapping can be, with lateral resolution down to the AFM tip radius and depth resolution even further enhanced due to the oblique measurement angle.

**Table I: Nanoscale UFM variability of the three regions (amorphous, transition, crystalline) for each material, image size, and surface. 'UFM variability' refers to the root mean square of the ultrasonic response in volts normalised to the ultrasonic amplitude as in ref [43], providing a more accurate means for comparing the ultrasonic response of measurements taken from different areas. Note: all BEXP images were acquired with 1 μm image size, except for the GST amorphous region (800 nm).**

| Material | Region | Flat Surface | BEXP Surface |
|---|---|---|---|
| $Ge_2Sb_2Te_5$ | Amorphous | 6.94 | 9.99 |
| $Ge_2Sb_2Te_5$ | Transition | 14.24 | 20.06 |
| $Ge_2Sb_2Te_5$ | Crystalline | 11.36 | 13.40 |
| GeTe | Amorphous | 3.59 | 3.92 |
| GeTe | Transition | 1.90 | 18.30 |
| GeTe | Crystalline | 4.01 | 17.78 |

Taking into account these possible artefacts and processing features, several insights into the nanomechanical properties of the three distinct phase regions can be made for both chalcogenide compositions. The amorphous regions for both GST and GT display the most uniform ultrasonic response, with a variation of just 10% and 4%, respectively. The crystalline and transition regions for GST and GT, on the other hand, display a much stronger nanomechanical contrast. For GST this variation is 13% and 20% for crystalline and transition regions, respectively, as compared to 18% and 18% for GT. The fact that the nanomechanical contrast varies most for the transition regions as compared to the purely crystalline or amorphous confirms the coexistence of nanoscale amorphous and crystalline phases within the transition region.

## 4.0 Conclusion

The nanomechanical morphology of amorphous and crystalline phases in $Ge_2Sb_2Te_5$ and GeTe thin films have been investigated by ultrasonic force microscopy, both in plan-view as well as for shallow-angle-cross sections prepared by beam exit Ar ion polishing. The characteristic length scale of the crystalline phase within the amorphous films is on the order of 100 – 200 nm for $Ge_2Sb_2Te_5$, consistent with the previously reported nucleation dominated crystallization behavior. Contrast in the nanomechanical response due to stiffness variations between the amorphous and crystalline phases are demonstrated up to 14% and 20% for the normal and cross-sectioned films, respectively. Several artefacts present in images of the cross-sectioned films were analyzed, with proposed suggestions for future sample fabrication and processing to improve similar measurements in the future. The advantages of utilizing ultrasonic force microscopy and beam exit Ar ion polishing are proven to be effective in characterizing materials with subtle variations in stiffness, relevant to the improvement of phase change films for data storage applications but also applicable to a much wider range of investigations into stiff materials with nanoscale heterogeneities.

## 5.0 Acknowledgements

JB and BDH recognize DOE, Basic Energy Sciences, Electron and Scanning Probe Microscopies, grant DE-SC0005037 for support. OVK and IG acknowledge support from the EPSRC grants EP/G06556X/1, EP/K023373/1 and EU grants QUANTIHEAT and FUNPROB.

# 6.0 References

[1] N. Yamada, E. Ohno, N. Akahira, K. Nishiuchi, K. Nagata, M. Takao, High-speed overwritable phase-change optical disk material, Jpn. J. Appl. Phys. Part 1 - Regul. Pap. Short Notes Rev. Pap., 26 (1987) 61-66.
[2] J. Akola, R.O. Jones, Structural phase transitions on the nanoscale: The crucial pattern in the phase-change materials Ge2Sb2Te5 and GeTe, Physical Review B, 76 (2007).
[3] S.R. Ovshinsky, Reversible Electrical Switching Phenomena in Disordered Structures, Physical Review Letters, 21 (1968) 1450-1453.
[4] W. Welnic, A. Pamungkas, R. Detemple, C. Steimer, S. Blugel, M. Wuttig, Unravelling the interplay of local structure and physical properties in phase-change materials, Nat. Mater., 5 (2006) 56-62.
[5] L. Goux, D.T. Castro, G.A.M. Hurkx, J.G. Lisoni, R. Delhougne, D.J. Gravesteijn, K. Attenborough, D.J. Wouters, Degradation of the Reset Switching During Endurance Testing of a Phase-Change Line Cell, Ieee T Electron Dev, 56 (2009) 354-358.
[6] T. Nonaka, G. Ohbayashi, Y. Toriumi, Y. Mori, H. Hashimoto, Crystal structure of GeTe and Ge2Sb2Te5 meta-stable phase, Thin Solid Films, 370 (2000) 258-261.
[7] S. Raoux, C.T. Rettner, J.L. Jordan-Sweet, A.J. Kellock, T. Topuria, P.M. Rice, D.C. Miller, Direct observation of amorphous to crystalline phase transitions in nanoparticle arrays of phase change materials, J. Appl. Phys., 102 (2007).
[8] J. Kim, M.H. Kwon, K.B. Song, Characterization of nanoscale recording mark on Ge2Sb2Te5 film, Ultramicroscopy, 108 (2008) 1246-1250.
[9] H.S.P. Wong, S. Raoux, S. Kim, J.L. Liang, J.P. Reifenberg, B. Rajendran, M. Asheghi, K.E. Goodson, Phase Change Memory, Proc. IEEE, 98 (2010) 2201-2227.
[10] S.J. Park, M.H. Jang, S.-J. Park, M.-H. Cho, D.-H. Ko, Characteristics of phase transition and separation in a In–Ge–Sb–Te system, Applied Surface Science, 258 (2012) 9786-9791.
[11] I. Grishin, B.D. Huey, O.V. Kolosov, Three-Dimensional Nanomechanical Mapping of Amorphous and Crystalline Phase Transitions in Phase-Change Materials, ACS Appl. Mater. Interfaces, 5 (2013) 11441-11445.
[12] W. Welnic, J.A. Kalb, D. Wamwangi, C. Steimer, M. Wuttig, Phase change materials: From structures to kinetics, Journal of Materials Research, 22 (2007) 2368-2375.
[13] S. Caravati, M. Bernasconi, T.D. Kuhne, M. Krack, M. Parrinello, Unravelling the Mechanism of Pressure Induced Amorphization of Phase Change Materials, Physical Review Letters, 102 (2009).
[14] Y. Fujisaki, Review of Emerging New Solid-State Non-Volatile Memories, Jpn. J. Appl. Phys., 52 (2013).
[15] F. Yang, L. Xu, R. Zhang, L. Geng, L. Tong, J. Xu, W.N. Su, Y. Yu, Z.Y. Ma, K.J. Chen, Direct observation of phase transition of GeSbTe thin films by Atomic Force Microscope, Applied Surface Science, 258 (2012) 9751-9755.
[16] J. Kim, Nanoscale Crystallization of Phase Change Ge2Sb2Te5 Film with AFM Lithography, Scanning, 32 (2010) 320-326.
[17] T. Gotoh, K. Sugawara, K. Tanaka, Minimal phase-change marks produced in amorphous Ge2Sb2Te5 films, Jpn J Appl Phys 2, 43 (2004) L818-L821.
[18] B.D. Huey, R.N. Premnath, S. Lee, N.A. Polomoff, High Speed SPM Applied for Direct Nanoscale Mapping of the Influence of Defects on Ferroelectric Switching Dynamics, Journal of the American Ceramic Society, 95 (2012) 1147-1162.
[19] A. Gyanathan, Y.C. Yeo, Multi-level phase change memory devices with Ge2Sb2Te5 layers separated by a thermal insulating Ta2O5 barrier layer, J. Appl. Phys., 110 (2011).

[20] K. Do, D. Lee, D.H. Ko, H. Sohn, M.H. Cho, TEM Study on Volume Changes and Void Formation in Ge2Sb2Te5 Films, with Repeated Phase Changes, Electrochemical and Solid State Letters, 13 (2010) H284-H286.
[21] O. V. Kolosov, I. Grishin, R. Jones, Material sensitive scanning probe microscopy of subsurface semiconductor nanostructures via beam exit Ar ionpolishing, IOP Nanotechnology, 22 (2011).
[22] A.P. McGuigan, B.D. Huey, G.A.D. Briggs, O.V. Kolosov, Y. Tsukahara, M. Yanaka, Measurement of debonding in cracked nanocomposite films by ultrasonic force microscopy, Appl. Phys. Lett., 80 (2002) 1180-1182.
[23] O.V. Kolosov, I. Grishin, R. Jones, Material sensitive scanning probe microscopy of subsurface semiconductor nanostructures via beam exit Ar ion polishing, Nanotechnology, 22 (2011).
[24] P. Sigmund, Sputtering by ion bombardment theoretical concepts, in: R. Behrisch (Ed.) Sputtering by Particle Bombardment I, Springer Berlin Heidelberg, 1981, pp. 9-17.
[25] F. Dinelli, M.R. Castell, D.A. Ritchie, N.J. Mason, G.A.D. Briggs, O.V. Kolosov, Mapping surface elastic properties of stiff and compliant materials on the nanoscale using ultrasonic force microscopy, Philos Mag A, 80 (2000) 2299-2323.
[26] M.T. Cuberes, G.A.D. Briggs, O. Kolosov, Nonlinear detection of ultrasonic vibration of AFM cantilevers in and out of contact with the sample, Nanotechnology, 12 (2001) 53-59.
[27] K. Yamanaka, S. Nakano, Ultrasonic atomic force microscope with overtone excitation of cantilever, Jpn. J. Appl. Phys. Part 1 - Regul. Pap. Short Notes Rev. Pap., 35 (1996) 3787-3792.
[28] K. Yamanaka, H. Ogiso, O. Kolosov, Ultrasonic force microscopy for nanometer resolution subsurface imaging, Appl. Phys. Lett., 64 (1994) 178-180.
[29] M. Libera, M. Chen, Time-Resolved reflection and transmission studies of amorphous ge-te thin-film crystallization, J. Appl. Phys., 73 (1993) 2272-2282.
[30] J.L. Bosse, P.D. Tovee, B.D. Huey, O.V. Kolosov, Physical mechanisms of megahertz vibrations and nonlinear detection in ultrasonic force and related microscopies, J. Appl. Phys., 115 (2014) -.
[31] R. Mitra, Effect of diameter-to-thickness ratio of crystal disks on the vibrational characteristics of ultrasonic ceramic transducers, Appl. Acoust., 48 (1996) 1-13.
[32] W.K. Njoroge, H.W. Woltgens, M. Wuttig, Density changes upon crystallization of Ge2Sb2.04Te4.74 films, J. Vac. Sci. Technol. A-Vac. Surf. Films, 20 (2002) 230-233.
[33] O.V. Kolosov, M.R. Castell, C.D. Marsh, G.A.D. Briggs, T.I. Kamins, R.S. Williams, Imaging the elastic nanostructure of Ge islands by ultrasonic force microscopy, Physical Review Letters, 81 (1998) 1046-1049.
[34] Y. Choi, Y.K. Lee, Elastic Modulus of Amorphous Ge2Sb2Te5 Thin Film Measured by Uniaxial Microtensile Test, Electron Mater Lett, 6 (2010) 23-26.
[35] I.M. Park, J.K. Jung, S.O. Ryu, K.J. Choi, B.G. Yu, Y.B. Park, S.M. Han, Y.C. Joo, Thermomechanical properties and mechanical stresses of Ge2Sb2Te5 films in phase-change random access memory, Thin Solid Films, 517 (2008) 848-852.
[36] J. Kalb, F. Spaepen, T.P.L. Pedersen, M. Wuttig, Viscosity and elastic constants of thin films of amorphous Te alloys used for optical data storage, J. Appl. Phys., 94 (2003) 4908-4912.
[37] Y. Won, J. Lee, M. Asheghi, T.W. Kenny, K.E. Goodson, Phase and thickness dependent modulus of Ge2Sb2Te5 films down to 25 nm thickness, Appl. Phys. Lett., 100 (2012).
[38] J.W. Hutchinson, Stresses and Failure Modes in Thin Films and Multilayers, in, Harvard University, Division of Engineering and Applied Sciences, 1996, pp. 4-7.
[39] H. Seo, T.H. Jeong, J.W. Park, C. Yeon, S.J. Kim, S.Y. Kim, Investigation of crystallization behavior of sputter-deposited nitrogen-doped amorphous Ge2Sb2Te5 thin films, Jpn. J. Appl. Phys. Part 1 - Regul. Pap. Short Notes Rev. Pap., 39 (2000) 745-751.

[40] J. Lee, Z.J. Li, J.P. Reifenberg, S. Lee, R. Sinclair, M. Asheghi, K.E. Goodson, Thermal conductivity anisotropy and grain structure in Ge2Sb2Te5 films, J. Appl. Phys., 109 (2011).

[41] Z.J. Li, J. Lee, J.P. Reifenberg, M. Asheghi, R.G.D. Jeyasingh, H.S.P. Wong, K.E. Goodson, Grain Boundaries, Phase Impurities, and Anisotropic Thermal Conduction in Phase-Change Memory, Ieee Electr Device L, 32 (2011) 961-963.

[42] F. Dinelli, H.E. Assender, N. Takeda, G.A.D. Briggs, O.V. Kolosov, Elastic mapping of heterogeneous nanostructures with ultrasonic force microscopy (UFM), Surf. Interface Anal., 27 (1999) 562-567.

[43] J.L. Bosse, P.D. Tovee, B.D. Huey, O.V. Kolosov, Physical mechanisms of megahertz vibrations and nonlinear detection in ultrasonic force and related microscopies, Journal of Applied Physics, 115 (2014) 144304.